\documentclass[a4paper,12pt]{article}
\usepackage[utf8]{inputenc}
\usepackage[margin=1in]{geometry}
\usepackage[usenames, svgnames]{xcolor}
\usepackage{setspace}
\usepackage{bbm}
\usepackage{graphicx}
\usepackage{multirow}
\usepackage{mathtools}
\usepackage{amsmath}
\usepackage{amssymb}
\usepackage{amsthm}
\usepackage{commath}
\usepackage{appendix}
\usepackage{enumitem}
\usepackage{float}
\usepackage{accents}
\usepackage{authblk}
\usepackage[colorlinks=true, citecolor=black]{hyperref}
\usepackage{leftidx}
\usepackage{natbib}
\usepackage{txfonts}
\usepackage{mathptmx}
\usepackage{tikz} %for drawing lines
\usepackage{url}
\usepackage[flushleft]{threeparttable} 
\usepackage{booktabs,caption}
\usepackage{rotating}
\usepackage{ulem}

\newcommand\dlim{\mathrel{\stackrel{\makebox[0pt]{\mbox{\normalfont\tiny d}}}{\longrightarrow}}}
\newcommand\plim{\mathrel{\stackrel{\makebox[0pt]{\mbox{\normalfont\tiny p}}}{\longrightarrow}}}

\newcommand{\argmin}{\operatornamewithlimits{arg\,min}}
\newcommand\mc[1]{\multicolumn{1}{c}{#1}}

\DeclareMathOperator{\E}{E}

\DeclareRobustCommand\dotted{\tikz[baseline=-0.6ex]\draw[thick,dotted] (0,0)--(0.54,0);}
\DeclareRobustCommand\dashed{\tikz[baseline=-0.6ex]\draw[thick,dashed] (0,0)--(0.54,0);}

\theoremstyle{definition}

\newtheorem{theorem}{Theorem}
\newtheorem{corollary}[theorem]{Corollary}
\newtheorem{lemma}{Lemma}
\newtheorem{proposition}{Proposition}

\newtheorem{assumption}{Assumption}

\providecommand{\keywords}[1]{\small	\textbf{Keywords:} #1}

\title{Causal Inference for Quantile Treatment Effects}

\author[1]{Shuo Sun}
\author[1]{Erica E. M. Moodie}
\author[2]{Johanna G. Ne{\v s}lehov{\'a}}
\date{}
\affil[1]{Department of Epidemiology, Biostatistics and Occupational Health, \newline McGill University, Montr\'eal, QC, Canada}
\affil[2]{Department of Mathematics and Statistics, McGill University, Montr\'eal, QC, Canada}
\begin{document}

\maketitle
\begin{abstract}
Analyses of environmental phenomena often are concerned with understanding unlikely events such as floods, heatwaves, droughts or high concentrations of pollutants. Yet the majority of the causal inference literature has focused on modelling means, rather than (possibly high) quantiles. We define a general estimator of the population quantile treatment (or exposure) effects (QTE) -- the weighted QTE (WQTE) -- of which the population QTE is a special case, along with a general class of balancing weights incorporating the propensity score. Asymptotic properties of the proposed WQTE estimators are derived. We further propose and compare propensity score regression and two weighted methods based on these balancing weights to understand the causal effect of an exposure on  quantiles, allowing for the exposure to be binary, discrete or continuous. Finite sample behavior of the three estimators is studied in simulation. The proposed methods are applied to data taken from the Bavarian Danube catchment area to estimate the 95\% QTE of phosphorus on copper concentration in the river.
\end{abstract}

\keywords{quantile treatment effect, weighted quantile treatment effect, propensity score,  quantile regression}

\thispagestyle{empty}

\section{Introduction}\label{sec1}
Environmental disturbances can lead to changes in the frequency or intensity of unusual environmental events such as floods, droughts, wildfires, hurricanes, and air pollution. Heuristically, an unlikely event can be seen as the occurrence of a value of an environmental variable that is above (or below) a threshold near the right (or left) tail of its historical distribution. Learning about the distributional causal effect beyond the average impact of specific interventions on environmental outcomes is often important for planning and policy-making. For example, chronic exposure to high levels of heavy metal can result in several consequences in the human body, such as liver and kidneys damage, leading to gradually progression of physical, muscular, and neurological degenerative processes that imitate diseases such as Parkinson's disease and Alzheimer's disease \citep{jaishankar2014toxicity}. Heavy metal pollution  often correlates with eutrophication in aquatic ecosystems \citep{lopez2003comparison}, where agricultural activity (e.g., fertilizers) is one of the major sources of phosphorus -- an element directly associated with eutrophication -- in aquatic ecosystems \citep{carpenter1998nonpoint,ekholm2000relationship}. Thus a policy-maker might be interested in the causal effect of a potential source of eutrophication (e.g., phosphorus) on heavy mental concentration on the tail (or some high quantile) of the underlying distribution.

In many areas of environmental science research, conventional methods based on correlation and regression are common data-based tools to analyze relationships. However, such approaches provide few insights into the causal mechanisms which might help discover and quantify the causal interdependence. In order to study causal relationships, causal inference techniques are needed; due to the increasing availability of observational and simulation data, there is a growing interest in these techniques in the environmental sciences literature.
For example, \cite{kretschmer2016using} investigated possible Arctic mechanisms which could be pivotal to understand northern hemisphere mid-latitude extreme winters in Eurasia and North America. Arctic teleconnection patterns are much less understood than their tropical counterparts and different climate models  provide partially conflicting results, thus the data-driven causal analyses are especially important \citep{shepherd2016effects,screen2018consistency,runge2019inferring}. Moreover, causal estimands have clear scientific interpretability and thus are better suited to inform policy-relevant decisions than statistical estimands based on conventional models. 
\cite{baccini2017assessing} assessed the short term causal impact of air pollution on health using a matching procedure based on the propensity score. \cite{samarasinghe2019study} investigated causal links between Arctic temperatures and  jet streams based on the concepts of Granger causality and Pearl causality. \cite{weisskopf2015air} used directed acyclic graphs (DAGs) to argue that the air pollution-autism association was not confounded by time-invariant factors and concluded that the causal association between air pollution and risk of autism was increasingly compelling. In the particular context of extreme event attribution, \cite{kirchmeier2017attributing} have shown that increased wildfire risk has been causally linked to anthropogenic forcings. There is also a growing literature on causal inference for environmental extremes (see \cite{asadi2015extremes,engelke2020graphical,mhalla2020causal,naveau2018revising}).

The focus in the classical causal inference literature has primarily been on the average treatment effect (ATE), that is, how exposures or treatments (e.g., a policy intervention in economics or a therapeutic agent in medicine) affect selected outcomes of interest; going forward, we shall use the terms ``treatment'' and ``exposure'' interchangeably.  However, low-probability large-impact environmental events (e.g., wildfires, air population) focus interest in the marginal causal effects in the tails rather than the mean. While an intervention that affects the tail of the distribution of an outcome of interest may also result in changes in the mean, shifts in the tail itself may be of greater importance. In this article, we thus focus on the so-called quantile treatment effect (QTE) \citep{doksum1974empirical}, which can characterize the potentially heterogeneous causal effect of an exposure on different points of the entire outcome distribution. There are two types of QTE, the conditional QTE (C-QTE), which analyzes the causal effect conditionally on confounding variables, and the population QTE, which represents the difference in the quantile of the outcome distribution that would be observed under the hypothetical setting in which the entire population received the exposure versus the entire population received some fixed alternative, thus marginalizing over the distribution of the confounders. In environmental research, population-level summaries are typically more desirable, both as these are more relevant for policy decisions, and as the relative effects at the individual level tend to be either small, or having large potential confounding bias difficult to rule out. Our aim is thus to develop inference for the population QTE. Contrary to the ATE, the population QTE is not easily recovered from the C-QTE as integrating the latter over the covariates typically does not yield the population QTE due to Jensen's inequality.

In previous literature, \cite{abadie2002instrumental} and \cite{chernozhukov2005iv} identified the conditional QTE using instrumental variables, but it is unclear whether this approach can be used to recover the population QTE. \cite{firpo2007efficient} suggested a semiparametric propensity score (PS) weighting estimator of population QTE, which was developed only for a binary exposure. \cite{cattaneo2010efficient} and \cite{cattaneo2013estimation} followed the work of \cite{hirano2003efficient} and \cite{firpo2007efficient} for binary ATEs and QTEs and extended inverse probability weighting (IPW) to a multivalued exposure context. One of the principle advantages of the propensity score is that it provides balance over a possibly high-dimensional vector of covariates while  adjusting or weighting based only on a scalar value. Nevertheless, IPW may perform poorly in practice when the exposure groups are initially quite different and when there exist extreme propensity scores (i.e., if estimated PSs are close to 0 or 1) \citep{lee2011weight,stuart2010matching}. Overlap weights (OW) have been proposed to mitigate the problems of bias and high variability observed in IPW \citep{li2018balancing,li2018addressing}. Overlap weights are bounded, smoothly reduce the influence of observations located in the tails of PS distribution, and minimize the asymptotic variance of the weighted ATE among a class of balancing weights. The application of overlap weights to estimate the population QTE has not been explored in the literature to date.

In this article, We define a more general formula for the population QTE -- the so-called weighted population QTE (WQTE) -- of which the population QTE is a special case. This leads us to propose a novel weighted QTE estimator, which can directly estimate the weighted population QTE by a simple two-step approach. We use a general class of balancing weights incorporating the PS to weight each group to a selected target population. Large sample properties of the estimator based on these weights are derived. We demonstrate that for a linear causal quantile model, the population QTE coincides with  C-QTE for causal quantile models without interactions between treatment and covariates (i.e., homogeneous C-QTE) and show that they differ when the C-QTE is heterogeneous.  We focus on three methods of estimating the population QTE, each based on variations of quantile regression: propensity score (PS) regression, inverse probability weighting (IPW), and overlap weighting (OW). The latter two weighted approaches are special cases of the WQTE estimator. When the C-QTE is heterogeneous, an empirical estimator based on PS regression is suggested. Accounting for homogeneous or heterogeneous conditional QTE, the consistency property of estimators corresponding to the two weighted methods is discussed and theoretically proven. We further extend the methods to allow the exposure to be binary, categorical or continuous. Despite the theoretical advantages of IPW and OW, their relative performance in a high quantile regression setting has not been studied. We thus conduct a series of simulation studies to compare these competing approaches of PS regression, IPW, and OW. 

This article is organized as follows. Section \ref{methodology} introduces assumptions, identifies the QTE, C-QTE and WQTE, presents properties of the nonparametric weighting estimators, and proposes the modelling framework for a  binary exposure. Section \ref{section3} extend the methods to categorical and continuous exposure settings. Section \ref{simulation} illustrates and compares the proposed methods through a series of simulations. In Section \ref{example}, we apply our methods to data from the Bavarian Danube River. We conclude in Section \ref{conclusion}. An online appendix with proofs of all results presented in this paper and additional material is available.

%=========================================
%             Methodology    
%=========================================
\section{Methodology} \label{methodology}
To measure the distributional (quantile) causal effect, this article will work within  the well-known Rubin causal model \citep{ rubin1978bayesian,rosenbaum1983central,holland1986statistics}. We first focus the discussion on the case where the exposure variable $Z$ takes two values, $Z=0$ and $Z=1$, as it simplifies the discussion.  The key models, parameter estimators,  and the  main conceptual issues  needed for implementation do not rely on this binary exposure simplification, and so in subsequent sections, we  generalize the methods to categorical and continuous exposures.

In a binary exposure setting, researchers can conceptualize exposure effects by considering the outcomes in alternative settings; these are known as \textit{potential outcomes}. Since a given unit can only be observed under one of the exposure alternatives, one of the potential outcomes is always missing while the other corresponds to the observed outcome. For unit $i$, let $Y_i$ denote the observed outcome, and let $Y_i(1)$ and $Y_i(0)$ be respectively the two potential outcomes under the state of being exposed and not exposed. As noted above, the potential outcomes are only partially observed because a given unit is either exposed or not. That is, if $Z_i=1$, we observe $Y_i(1)$; otherwise if $Z_i=0$, we observe $Y_i(0)$.

As noted by \cite{greenland1999confounding}, counterfactual (or potential outcomes) approaches to causal inference emphasize the importance of randomization \citep{rubin1978bayesian,rubin1990comment,rubin1991practical,greenland1986identifiability, robins1986new,greenland1990randomization}. However, in observational studies, no such assurance is available and confounding is present when the assumption $(Y(0), Y(1))\perp Z$ does not hold. It may be that $(Y(0), Y(1))$ and $Z$ have common determinants. These determinants, the so-called confounders, are causes of both the exposure and the outcome. Conditional on these confounders, $(Y(0), Y(1))$ are independent of exposure variable $Z$; of course, for this to be useful in practice, all such confounders must be measured (without error).

\subsection{Main assumptions}
Assume that $\boldsymbol{X} = (X_1,...,X_p)^\top$ is a vector of observed pre-exposure covariates of an outcome $Y$ with exposure variable $Z$. The propensity score is given by $e(\boldsymbol{X}) = \Pr\{Z = 1\mid \boldsymbol{X}\}$.  In order to ensure a consistent estimator of a causal effect, we must make the following assumptions \citep{rubin1978bayesian, rubin1980randomization,rosenbaum1983central}:
\begin{assumption} The following conditions hold: \label{assumption:1}
\end{assumption}
\begin{enumerate}[label=(\roman*)]
    \item \textit{Consistency}: The observed outcome is $Y = Z \cdot Y(1) + (1-Z)\cdot Y(0)$;
    \item \textit{Unconfoundedness}: $(Y(0), Y(1))\perp Z \mid \boldsymbol{X}$;
    \item \textit{Positivity}:  For some $c>0$, $c < e(\boldsymbol{x}) < 1-c$, for all $\boldsymbol{x} \in \mathcal{X} \subset \mathbbm{R}^p$, where $\mathcal{X}$ is the support of $\boldsymbol{X}$;
    \item \textit{No interference}: One individual’s outcome is not affected by whether another individual was exposed.
\end{enumerate}

\subsection{Identification of the QTE and C-QTE}
Let $F_{Y(1)}$ and $F_{Y(0)}$, respectively, be the unconditional CDFs of the potential outcomes $Y(1)$ and $Y(0)$. \cite{doksum1974empirical} shows that for all $y$, if letting $\Delta(y)$ be a smallest value such that $F_{Y(0)}(y) = F_{Y(1)}(y+\Delta(y))$, then $\Delta(y) = F^{-1}_{Y(1)}(F_{Y(0)}(y)) - y$. We shall make the additional assumption  that for values $\tau \in (0,1)$, the CDFs of potential outcomes are continuous and strictly increasing so that these inverse functions are well defined. Thus, for $\tau = F_{Y(0)}(y)$, the $\tau$th population quantile treatment effect is given by 
\begin{align}\label{meth:eq:1}
    \Delta_{\text{QTE}}(\tau) &= \Delta(y) = F^{-1}_{Y(1)}(\tau) - F^{-1}_{Y(0)}(\tau), 
\end{align}
the difference between the quantiles of the two marginal distributions \citep{koenker1978regression}, with
\begin{equation}
\begin{aligned} 
    F_{Y(j)}(y) &= \int F_{Y(j)\mid \boldsymbol{X}}(y\mid \boldsymbol{x}) f(\boldsymbol{x})d\boldsymbol{x}, \quad j=0, 1,
\end{aligned}
\label{meth:qte}    
\end{equation}
assuming that the marginal density of the pre-exposure covariates $\boldsymbol{X}$, $f(\boldsymbol{X})$, exists, with respect to a base measure which is a product of counting measures corresponding to categorical variables and the Lebesgue measure corresponding to continuous variables. Intuitively, $\Delta_{\text{QTE}}(\tau)$ may be thought of as the ``horizontal distance'' between $F_{Y(1)}$ and $F_{Y(0)}$ \citep{doksum1974empirical}. For a given $\tau \in (0,1)$, 
let $\Delta(y,\boldsymbol{x})$ denote the $\tau$th C-QTE given the covariates  $\boldsymbol{X}=\boldsymbol{x}$, i.e., a smallest value such that for all $\boldsymbol{x} \in \mathcal{X}$,
$F_{Y(0)\mid\boldsymbol{X}}(y\mid\boldsymbol{x}) = F_{Y(1)\mid \boldsymbol{X}}(y+\Delta(y, \boldsymbol{x})\mid \boldsymbol{x})$. The $\tau$th conditional quantile treatment effect is given by 
\begin{align} \label{def:c-qte}
    \Delta_{\text{C-QTE}}(\tau) &= \Delta(y,\boldsymbol{x}) = F^{-1}_{Y(1)\mid\boldsymbol{X}}(\tau\mid\boldsymbol{x}) - F^{-1}_{Y(0)\mid\boldsymbol{X}}(\tau\mid\boldsymbol{x}).
\end{align}
We shall also make the additional assumption that for values $\tau \in (0,1)$, the conditional CDFs of potential outcomes are continuous and strictly increasing.

\subsection{Model Framework}%\label{modelframe}
Define the $\tau$th conditional linear quantile regression of $y$ for given treatment $Z=z$, and covariates $\boldsymbol{x} = (x_1,...,x_p)^\top$ as
\begin{align}
    F^{-1}_{Y\mid X, Z}(\tau \mid  \boldsymbol{x}, z)&= \beta_0 + \beta_1 z + \boldsymbol{\beta}_{2} z \boldsymbol{x} + \boldsymbol{\beta}_{3}\boldsymbol{x} + F^{-1}_{\epsilon}(\tau),
    \label{mfram:loc_q}
\end{align}
where $\epsilon$ denotes the random error term. We assume here that the errors are i.i.d with a common CDF $F_{\epsilon}$ which is continuous and strictly increasing. The vectors  $\boldsymbol{\beta}_{2}$ and  $\boldsymbol{\beta}_{3}$ of parameters have dimension $p$. This model may be seen as arising from the linear model 
\begin{equation} \label{lm}
\begin{aligned}
y=\beta_0 + \beta_1 z + \boldsymbol{\beta}_{2} z \boldsymbol{x} + \boldsymbol{\beta}_{3}\boldsymbol{x} + \epsilon
\end{aligned}
\end{equation}
with i.i.d homogeneous errors. As the distribution of errors is independent of $Z$ and $\boldsymbol{X}$, the exposure and confounders affect only the location of the response $Y$. We will call model (\ref{mfram:loc_q}) the location shift model.

\begin{proposition} \label{pqte=cqte}
For the location shift model (\ref{mfram:loc_q}) under the restriction of a homogeneous exposure effect (i.e., $\boldsymbol{\beta}_2 = \boldsymbol{0}$),  the C-QTE given by $\beta_1$ coincides with population QTE, $\Delta_{\text{QTE}}(\tau)$, for any fixed $\tau$.
\end{proposition}
The proposition above is proved in the Supplementary Materials. It holds since the marginal CDFs of the exposed and unexposed groups only differ by a constant location shift $\beta_1$, such that $\beta_1$ is the population QTE. As proposed by \cite{koenker1978regression} and used here, estimators of $\boldsymbol{\beta} = (\beta_0, \beta_1, \boldsymbol{\beta}_{2}, \boldsymbol{\beta}_{3})^\top$ can be obtained by minimizing the sum of piece-wise linear check functions $\rho_{\tau}(u) = u(\tau-\textit{I}(u < 0))$:
\begin{align}
    \hat{\boldsymbol{\beta}} &= \argmin_{\boldsymbol{\beta}}\sum \rho_{\tau}(y_i - F^{-1}_{Y\mid X, Z}(\tau \mid \boldsymbol{x_i}, z_i)).
    \label{mfram:eq:2}
\end{align}
Based on the location shift model (\ref{mfram:loc_q}), we first propose a propensity score regression approach to estimate the population QTE in the following subsection.

\subsection{Propensity Score Regression}
The true propensity score has what is known as the balancing property \citep{rosenbaum1983central}: $Z_i \perp \boldsymbol{X_i} \mid e(\boldsymbol{X_i})$. That is, conditional on the propensity score, the distribution of the multidimensional $\boldsymbol{X}$ is the same between exposed and unexposed groups. Under Assumption \ref{assumption:1}, the balancing property implies: $(Y_i(1), Y_i(0)) \perp Z_i \mid e(\boldsymbol{X_i})$. A model for propensity score quantile regression can be expressed as
\begin{align}\label{ps_reg}
    F^{-1}_{Y\mid e(\boldsymbol{X}), Z}(\tau \mid e(\boldsymbol{x}), z)&= \beta_0(\tau) + \beta_1 z + \beta_2ze(\boldsymbol{x})+ \beta_3e(\boldsymbol{x}). 
\end{align}   
In this representation, $\beta_0(\tau)$ absorbs $\beta_0$ from the linear model \eqref{mfram:loc_q} and the quantile function of the error $\epsilon$. Coefficient estimates are obtained using the minimization in Eq.~\eqref{mfram:eq:2}. Here, $\hat \beta_1$ is the estimated C-QTE. Following Proposition \ref{pqte=cqte}, we show in the Supplementary Materials that
\begin{proposition} \label{prop:ps_reg}
If $\beta_2=0$ in location shift PS regression model \eqref{ps_reg}, then the population QTE equals $\beta_1$, the C-QTE.
\end{proposition}
Proposition \ref{prop:ps_reg} implies that in the homogeneous QTE setting, the population QTE can be estimated directly using the conditional QTE estimate. However, when interactions are present, that is, when we believe the conditional QTE is not homogeneous across $\boldsymbol{X}$, this is no longer possible. Under hetergeneous QTE, we must integrate out the PS in order to recover the population QTE: 
\begin{equation} \label{recover:ps_reg}
\begin{aligned} 
    F_{Y(j)}(y) &= \int F_{Y(j)\mid e(\boldsymbol{X})}(y\mid e(\boldsymbol{X})=t) dF_{e(\boldsymbol{X})}(t) = \int F_{Y\mid e(\boldsymbol{X}), Z=j}(y\mid e(\boldsymbol{X})=t, Z=j) dF_{e(\boldsymbol{X})}(t), \quad j=0, 1. 
\end{aligned}
\end{equation}
Calculating the distribution of the PS is not easy. A possible empirical approximation is detailed in the Online Supplementary Materials and proceeds as follows. First, a sequence of quantile regressions $\hat F^{-1}_{Y\mid e(\boldsymbol{X}), Z}(\tau_i\mid e(\boldsymbol{X}),Z)$ with $(\tau_1,...,\tau_K) \in (0,1)$ is fitted. This allows us to calculate $\hat F_{Y\mid e(\boldsymbol{X}), Z}(y\mid e(\boldsymbol{X}), Z) = \frac{1}{K}\sum_{k=1}^K\mathbbm{1}\{\hat F^{-1}_{Y\mid e(\boldsymbol{X}), Z}(\tau_i\mid e(\boldsymbol{X}),Z)\leq y\}$. The marginal CDF can then be estimated as $\hat F_{Y(j)}(y) = \frac{1}{n}\sum_{i=1}^n \hat F_{Y\mid e(\boldsymbol{X}), Z}(y\mid e(\boldsymbol{X_i}), Z=j)$ for $j=0,1$. Upon repeating this procedure for a number of $y$ values from the support, the marginal quantiles can be estimated empirically by $\hat F_{Y(j)}^{-1}(\tau)= \inf\{y: \hat F_{Y(j)}(y)\geq \tau\}$. This empirical method is computationally costly. To circumvent this problem, we now propose a weighted method to compute the population QTE directly with certain balancing weights. 

\subsection{Identification of the WQTE}
As is typical in causal comparisons (e.g.~the ATE), some functions of potential outcomes are compared over a target population. Moreover, if there is insufficient overlap in the distribution of exposed and unexposed groups, one may wish to restrict the analysis to a sub-population so that there is sufficiently large probability of observing both groups \citep{hirano2003efficient}. Suppose the marginal density of $\boldsymbol{X}$, denoted as $f(\boldsymbol{X})$, exists. We represent a specific target population as $f(\boldsymbol{X})g(\boldsymbol{X})$, where $g(\cdot)$ is a pre-specified function of $\boldsymbol{X}$ determining the target population. In this article, we restrict $g$ to be positive, and such that $\E[g^2(\boldsymbol{X})]  < \infty$. We then define the $\tau$th weighted quantile treatment (exposure) effect as
\begin{align}\label{meth:eq:wqte1}
    \Delta_{\text{WQTE}}(\tau) &=  \leftidx{^w}F^{-1}_{Y(1)}(\tau) - \leftidx{^w}F^{-1}_{Y(0)}(\tau),
\end{align}
with
\begin{equation} \label{meth:eq:wqte2}
\begin{aligned}
    \leftidx{^w}F_{Y(j)}(y) &= \frac{\int F_{Y(j)\mid \boldsymbol{X}}(y\mid \boldsymbol{x}) f(\boldsymbol{x})g(\boldsymbol{x})d\boldsymbol{x}}{\int f(\boldsymbol{x})g(\boldsymbol{x})d\boldsymbol{x}}, \quad j=0, 1.
\end{aligned}
\end{equation}
The WQTE can be seen as a generalization of the weighted average treatment effect (WATE) discussed by \cite{hirano2003efficient}. The pre-specified function $g(\cdot)$ defines the target population and estimand via determining the weights. It allows to compare the potential outcomes over a subpopulation of interest. The population QTE is a special case of the WQTE that arises when $g(\boldsymbol{x}) = 1$. When $g(\boldsymbol{x})$ equals the propensity score $e(\boldsymbol{x})$, this leads to the QTE under exposed (analogous to the average treatment effect on the treated). When $g(\boldsymbol{x}) = \Pr\{Z=0\mid\boldsymbol{x}\} = 1-e(\boldsymbol{x})$, the target population is the unexposed subpopulation, the estimand is the population QTE for the unexposed. The choice of $g(\cdot)$ need not be a function of the propensity score, it can take any form, reflecting statistical and scientific considerations. For example, if age is the covariate and we are interested in estimating the QTE for people within a specific age range, we could define a $g(\cdot)$ to select this target population.

Recall that the C-QTE given $\boldsymbol{x}$ may be either homogeneous or heterogeneous; in the former setting,  $\Delta_{\text{QTE}}(\tau)$ coincides with $\Delta_{\text{WQTE}}(\tau)$ with homogeneous exposure effect. 
\begin{theorem} \label{qte_equal_thm}
If the $\tau$th C-QTE is homogeneous, that is, if $\Delta(y, \boldsymbol{x})=\Delta(y)$ for almost all $x \in \mathcal{X}$, then $\ \Delta_{\text{QTE}}(\tau)=\Delta_{\text{WQTE}}(\tau)$.
\end{theorem}
The proof of Theorem \ref{qte_equal_thm} is provided in the Online Supplementary Materials.

\subsection{Large Sample Properties}
Let $q_{1,\tau}=\leftidx{^w}{F^{-1}_{Y(1)}(\tau)}$, $q_{0,\tau}=\leftidx{^w}{F^{-1}_{Y(0)}(\tau)}$. Before showing the main theorems in this section, we first provide a lemma which states that the quantiles of the potential outcomes can be written as functions of observed outcomes.  

\begin{lemma} \label{lemma_obs}
Under Assumption \ref{assumption:1}, for all $\tau\in (0,1)$,
\begin{align*}
\tau &= \leftidx{^w}F_{Y(1)}(q_{1,\tau}) = \frac{1}{\E[g(\boldsymbol{X})]}\E\Big[\frac{g(\boldsymbol{X})Z}{e(\boldsymbol{X})}\mathbbm{1}\{Y\leq q_{1,\tau}\}\Big],\\
\tau &= \leftidx{^w}F_{Y(0)}(q_{0,\tau}) = \frac{1}{\E[g(\boldsymbol{X})]}\E\Big[\frac{g(\boldsymbol{X})(1-Z)}{1-e(\boldsymbol{X})}\mathbbm{1}\{Y\leq q_{0,\tau}\}\Big].
\end{align*}
\end{lemma}
We can now show that the sample WQTE estimator %, $hat{\Delta}_{\text{WQTE}}= \hat q_{1,\tau} - \hat q_{0,\tau}$,
\begin{align}\label{wqte_est1}
\hat{\Delta}_{\text{WQTE}}= \hat q_{1,\tau} - \hat q_{0,\tau},
\end{align}
obtained via solving a weighted sum of check functions
\begin{equation}
\begin{aligned}
\hat q_{1,\tau} &= \argmin_{q} \sum w_{1,i} \cdot \rho_{\tau}(y_i - q),\quad 
\hat q_{0,\tau} &= \argmin_{q} \sum  w_{0,i} \cdot \rho_{\tau}(y_i - q),
\end{aligned}
\label{q_est} 
\end{equation}
with weights given by 
\begin{equation}\label{g_ws}
w_{1,i}(\boldsymbol{x_i})=\frac{g(\boldsymbol{x}_i)}{e(\boldsymbol{x}_i)}z_i, \quad
w_{0,i}(\boldsymbol{x}_i) =\frac{g(\boldsymbol{x}_i)}{1- e(\boldsymbol{x}_i)}(1-z_i)
\end{equation}
is a consistent estimator of $\Delta_{\text{WQTE}}(\tau)$ and is asymptotically normal under  mild assumptions specified below. First assume that $Y(j)\mid\boldsymbol{X}$ has a density $f_{Y(j)\mid\boldsymbol{X}}$ for $j=0,1$. Define
\begin{equation*}
\begin{aligned}
&D_{j}(Y) = \mathbbm{1}\{Y\leq q_{j,\tau}\} -\tau,\quad j=0,1,\\
&\psi_{1,\tau}(Y,\boldsymbol{X},Z) =  \frac{g(\boldsymbol{X})Z}{e(\boldsymbol{X})}D_1(Y)-\frac{g(\boldsymbol{X})(Z-e(\boldsymbol{X}))}{e(\boldsymbol{X})}\E[D_1(Y)\mid \boldsymbol{X}, Z=1],\\
&\psi_{0,\tau}(Y,\boldsymbol{X},Z) =  \frac{g(\boldsymbol{X})(1-Z)}{1-e(\boldsymbol{X})}D_0(Y)+\frac{g(\boldsymbol{X})(Z-e(\boldsymbol{X}))}{1-e(\boldsymbol{X})}\E[D_0(Y)\mid \boldsymbol{X}, Z=0].\\
\end{aligned}
\end{equation*}
\begin{theorem}\label{thm_properties_tPS}
Suppose that Assumption \ref{assumption:1} holds.  Then for any fixed $\tau \in (0,1)$, when $n\rightarrow \infty$, 
\begin{enumerate}[label=(\roman*)]
    \item $\hat \Delta_{\text{WQTE}}(\tau) \plim  \Delta_{\text{WQTE}}(\tau)$, where $\plim$ denotes convergence in probability.
    \item $\sqrt{n}\bigr(\hat \Delta_{\text{WQTE}}(\tau)- \Delta_{\text{WQTE}}(\tau)\bigl)\dlim N(0, V_{\tau})$, where 
    $V_{\tau} = \E\Big[\Big(\frac{\psi_{0,\tau}(Y,\boldsymbol{X},Z)}{\E[f_{Y(0)\mid \boldsymbol{X}}(q_{0,\tau}\mid \boldsymbol{X})g(\boldsymbol{X})]}-\frac{\psi_{1,\tau}(Y,\boldsymbol{X},Z)}{\E[f_{Y(1)\mid \boldsymbol{X}}(q_{1,\tau}\mid \boldsymbol{X})g(\boldsymbol{X})]}\Big)^2\Big]$
    \item[] and $\dlim$ denotes convergence in distribution.
\end{enumerate}
\end{theorem}
In applications, the true PS $e(\boldsymbol{x})$ in Eq.~\eqref{g_ws} is unknown and is replaced by the estimated propensity score $\hat e(\boldsymbol{x})$. However, with a uniformly consistent estimator of PS, the consistency and asymptotic normality stated in Theorem \ref{thm_properties_tPS} still hold. We set the following restriction:
\begin{assumption}\label{assumption:ps_restiction}
The estimated propensity score $\hat e(\boldsymbol{X})$ satisfies the following conditions:
\begin{enumerate}[label=(\roman*)]
\item $\hat e(\boldsymbol{X})$ is uniformly consistent to the true propensity score $e(\boldsymbol{X})$;
\item  For each $\boldsymbol{x}\in\mathcal{X}$, $\hat e(\boldsymbol{x})$ is bounded away from zero and one: $0<\epsilon_1\leq \hat e(\boldsymbol{x}) \leq \epsilon_2 < 1$.
\end{enumerate}
\end{assumption}
Under these assumptions and with Lemma \ref{lemma_obs}, we can state the following result:
\begin{theorem}\label{thm_properties}
Suppose that the assumptions of Theorem~\ref{thm_properties_tPS} and Assumption \ref{assumption:ps_restiction} hold and that the  weights in Eq.~\eqref{q_est} and \eqref{g_ws} are replaced with 
\begin{equation}\label{eq:hatweights}
\hat w_{1,i}(\boldsymbol{x}_i)=\frac{g(\boldsymbol{x}_i)}{\hat e(\boldsymbol{x}_i)}z_i, \quad  
\hat w_{0,i}(\boldsymbol{x}_i) =\frac{g(\boldsymbol{x}_i)}{1- \hat e(\boldsymbol{x}_i)}(1-z_i).
\end{equation}
Then statements (i) and (ii) in 
Theorem \ref{thm_properties_tPS} remain valid.
\end{theorem}
The proof of Theorem~\ref{thm_properties} is provided in the Online Supplementary Materials. The weights defined in \eqref{g_ws} are balancing weights because they ensure that the weighted distributions of the covariates between comparison groups are the same (see \cite{li2018balancing}). In view of Theorems \ref{qte_equal_thm} and \ref{thm_properties_tPS}, we have the following result.
\begin{corollary} \label{consist_est_w}
Suppose that the assumptions of Theorem~\ref{thm_properties} hold and that for a given $\tau \in (0,1)$  the C-QTE is homogeneous. Then $\hat \Delta_{\text{WQTE}}(\tau)$ in \eqref{wqte_est1} with weights as in \eqref{eq:hatweights} is a  consistent and asymptotically normal estimator of $\Delta_{\text{QTE}}(\tau)$.
\end{corollary}
In the next two subsections, we consider estimators for two special cases of the WQTE.

\subsection{Inverse Probability Weighting}
Inverse probability weighted estimators were proposed by \cite{horvitz1952generalization} for surveys in which subjects are sampled with unequal probabilities. The idea behind weighting is that the distributions of the covariates $\boldsymbol{X}$ are imbalanced between exposed and unexposed groups (representing, potentially, confounding) and that balance can be achieved by reweighting. The quantile regression model is represented as
\begin{align}
    F^{-1}_{Y|Z}(\tau \mid z)&= \beta_0^*(\tau) + \beta_1^* z,  
    \label{mfram:eq:4}
\end{align}
and under weighted regression, $\hat \beta_1^*(\tau)$ is obtained by solving a weighted sum of check functions 
\begin{align}
    \hat{\boldsymbol{\beta}^*}(\tau) &= \argmin_{\boldsymbol{\beta}(\tau)}\sum_{i=1}^n w_i \cdot \rho_{\tau}(y_i - F^{-1}_{Y|Z}(\tau \mid z_i)). 
    \label{mfram:eq:5}
\end{align}
Letting $g(\boldsymbol x) =1$ in Eq.~(\ref{g_ws}), the inverse probability weights $w_i$ are given by
\begin{align}
    w_i &= \frac{z_i}{e(\boldsymbol{x_i})} + \frac{1-z_i}{1-e(\boldsymbol{x_i})}.
    \label{mfram:eq:3}
\end{align}
The IPW approach directly estimates the population QTE, which is $\beta_1^*$. With $g(\boldsymbol{x})=1$, WQTE estimator formulated in (\ref{meth:eq:wqte1}) and (\ref{meth:eq:wqte2}) simplifies to the population QTE. Therefore, the IPW estimator is also consistent for $\Delta_{\text{QTE}}(\tau)$ by Theorem~\ref{thm_properties_tPS}. Formally, we have
\begin{corollary} \label{ipw_consist}
Suppose that Assumption \ref{assumption:1} holds. Then the IPW estimator is a consistent estimator of the population QTE.
\end{corollary}
The proof for this theorem follows directly from the stated assumptions, the definition of WQTE and Theorem \ref{thm_properties_tPS}.

\subsection{Overlap Weighting}
Inverse probability weights in \eqref{mfram:eq:3} are positive but  are sensitive to extreme PS values. Extreme weights are likely to result in finite-sample bias and increased variance in the estimate of the causal parameter $\Delta_{\text{QTE}}(\tau)$. \cite{li2018balancing} proposed overlap weights, which are obtained by letting $g(\boldsymbol{x})=e(\boldsymbol{x})(1-e(\boldsymbol{x}))$ in Eq.~(\ref{g_ws}), viz. 
\begin{align}
    w_i &= z_i\cdot(1-e(\boldsymbol{x_i}))+(1-z_i)\cdot e(\boldsymbol{x_i}). \label{eq:7}
\end{align}
These weights were first introduced as `absolute value weights' in \cite{wallace2015doubly} as they can also be expressed as $w_i = |z_i-e(\boldsymbol{x_i})|$. The idea behind overlap weighting is that $g(\boldsymbol{x})=e(\boldsymbol{x})(1-e(\boldsymbol{x}))$ is maximized when PS is $0.5$, and decreases to $0$ when PS reaches extreme values of 0 or 1. By this construction, overlap weights upweight units with $e(\boldsymbol{x_i})$ close to $0.5$, relative to the units located in the tails of the distribution of PS \citep{li2018addressing}. We propose overlap weighted quantile regression,  conducted by replacing IPW in (\ref{mfram:eq:5}) with OW. As noted in Theorem \ref{thm_properties_tPS}, the OW estimator is consistent for $\Delta_{\text{WQTE}}(\tau)$. From Theorem \ref{qte_equal_thm}, under the condition that the conditional QTE is homogeneous (i.e., no interactions between exposure and covariates), the OW estimator is also a consistent estimator of population QTE $\Delta_{\text{QTE}}(\tau)$. Formally,
\begin{corollary} \label{ow_consist}
Suppose that Assumption \ref{assumption:1} holds. For any fixed $\tau$, if the C-QTE is homogeneous, the OW estimator is a consistent estimator of the population QTE.
\end{corollary}
The proof for Corollary~\ref{ow_consist} follows directly from Theorem \ref{qte_equal_thm} and Theorem \ref{thm_properties_tPS} by noting that the overlap weights are obtained when $g(\boldsymbol{x})=e(\boldsymbol{x})(1-e(\boldsymbol{x}))$. Note that by Corollary~\ref{ow_consist}, consistency of the overlap weighted estimator for the QTE requires homogeneous exposure effects in contrast to Corolarry \ref{ipw_consist}, where homogeneity is not required for consistency of the IPW estimator. As stated in Theorem \ref{thm_properties}, with a uniformly consistent estimate of the PS that is bounded away from zero and one, the consistency properties discussed in the two weighted approaches still hold. 

\section{Extension to Categorical and Continuous Exposure}\label{section3}
\subsection{Categorical Exposure}
In this section, we allow the exposure to take integer values between 1 and $J$, $J\geq 3$, so that $Z \in \{1, 2, ...,J\}$. \cite{imbens2000role} defines the generalized propensity score as the conditional probability of receiving a particular level of exposure $Z=j$, $j\in\{1,2,...,J\}$, given the pre-exposure covariate $\boldsymbol{X}$,
\begin{align}
    e_j(\boldsymbol{X}) &= \Pr(Z=j \mid \boldsymbol{X}), \label{eq:8}
\end{align}
where $\sum_j e_j(\boldsymbol{X}) = 1$. Then for unit $i$, if $Z_i = j$, $j\in\{1,2,...,J\}$, we observed $Y_i(j)$. A given unit only receives one level of treatment, thus $Y_i(1), Y_i(2),...,Y_i(J)$ are the potential outcomes associated with receiving different levels of exposure. The four points in Assumption \ref{assumption:1} for binary exposure can be generalized to the categorical exposure case. For example, the generalized unconfoundedness assumption is $\{Y(j) \perp Z\} \mid \boldsymbol{X}$, for all $j\in\{1,2,...,J\}$. In the categorical exposure case, we can estimate pairwise population QTEs with a chosen baseline. Without loss of generality, we choose $Z=1$ as the baseline or reference exposure level and construct PS regression, IPW and OW causal quantile models. Let $\boldsymbol{D} = (\mathbbm{1}(Z=2), ..., \mathbbm{1}(Z=J))^\top$. 

\textit{PS regression}: In the categorical exposure setting, the PS regression is defined by
\begin{align} \label{ps_reg_gen}
F^{-1}_{Y\mid e(\boldsymbol{X}),Z}(\tau \mid e(\boldsymbol{x}),z) = \beta_0(\tau) + \boldsymbol{\beta}_{1} \boldsymbol{D} + \boldsymbol{\beta}_{2}\boldsymbol{M} + \boldsymbol{\beta}_{3}\boldsymbol{e}(\boldsymbol{x}),
\end{align}
where $\boldsymbol{M}$ is a $(J-1)^2$ dimensional vector with elements $\mathbbm{1}(Z=j)\times e_k(\boldsymbol{x})$, for all $j,k \in \{2, 3, ...,J\}$, $\boldsymbol{e(x)} = (e_2(\boldsymbol{x}),...,e_J(\boldsymbol{x}))^\top$. The vectors $\boldsymbol{\beta}_{1}, \boldsymbol{\beta}_{2}, \boldsymbol{\beta}_{3}$ are coefficients  corresponding to $\boldsymbol{D}$, $\boldsymbol{M}$ and $\boldsymbol{e(x)}$, respectively; their estimates are obtained by solving the minimization in Eq.~\eqref{mfram:eq:2}. We generalize Proposition \ref{prop:ps_reg} to categorical exposure: if $\boldsymbol{\beta}_2 = \boldsymbol{0}$, the pairwise population QTEs are contained in $\boldsymbol{\beta}_1$. When homogeneity does not hold (i.e.~interactions exist), as before, it is necessary to marginalize over the propensity score distribution to recover the pairwise population QTEs, which in general cannot be done exactly.

\textit{IPW}: In the categorical exposure setting, the outcome regression is defined by
\begin{equation}
\begin{aligned}
F^{-1}_{Y\mid Z}(\tau \mid z) = \beta_0^*(\tau) + \boldsymbol{\beta}_{1}^* \boldsymbol{D}
\end{aligned}
\end{equation}
where an estimate of $\boldsymbol{\beta}_{1}^*$ is obtained by solving the optimization problem in Eq.~(\ref{mfram:eq:5}) with $w_i = 1/e_{z_i}(\boldsymbol{x_i}) = 1/\Pr(Z_i=z_i\mid\boldsymbol{x_i})$, with $z_i$ being the actual exposure received by unit $i$.

\textit{OW}: Define the generalized overlap weights as $w_i = (1/e_{z_i}(\boldsymbol{x_i}))(\sum_{j=1}^J1/e_j(\boldsymbol{x_i}))^{-1}$, with $z_i$ being the actual exposure received by unit $i$ \citep{li2019propensity}. For binary exposure with $J=2$, the generalized overlap weights reduce to the overlap weights proposed by \cite{li2018balancing}.

\subsection{Extension to Continuous Exposure}
We can further generalize to continuous exposure scenario. Let $\mathcal{Z}$ denotes the support of exposure values $Z$, where $Z$ is continuously distributed on $\mathcal{Z}$. Let $f_{Z\mid\boldsymbol{X}}$ be the conditional density of exposure $Z$ given observed confounding $\boldsymbol{X}$. The generalized propensity score in continuous exposure setting \citep{imbens2000role} is defined as 
\begin{align*}
    e_z(\boldsymbol{X}) &= f_{Z|\boldsymbol{X}}(z \mid \boldsymbol{X}). 
\end{align*}
When confounders $\boldsymbol{X}$ are strongly associated with $Z$, these probabilities may vary greatly between subjects. This variability can result in extremely large values of the inverse probability score weights for some subjects \citep{robins2000marginal}, and care must be taken to verify that positivity has not been violated. 

Following \cite{naimi2014constructing}, we consider binning the continuous exposure values according to its empirical quantiles and treating each bin as a categorical exposure variable. The resulting weights can be used to approximately adjust for confounding, while keeping exposure in its continuous form in the outcome model. Let $\{z_1, z_2, ..., z_m\}$ denote the categorized exposure levels based on $m$ bins. In particular, let $q_k$, $k\in\{0,1,2, ..., m\}$ denote the cut-points for these $m$ bins.  The generalized PS can then be computed by $e_{z_j}(\boldsymbol{X}) = F(q_j\mid \boldsymbol{X}) - F(q_{j-1}\mid \boldsymbol{X})$, where $F(\cdot\mid\boldsymbol{X})$ was the true conditional distribution of continuous exposure $Z$. Alternatively, if the true distribution of the treatment is unknown, the generalized PS can be estimated via multinomial regression on the $m$ constructed bins of exposure; this approach is a practical and feasible approach to avoid strong requirements on knowledge of the exposure distribution. Following generalized assumptions of Assumption \ref{assumption:1}, the three causal quantile approaches can be again be considered: PS regression, IPW, and  OW, with implementation much as in the categorical exposure setting.

\section{Simulation} \label{simulation}
In the previous section, three methods for estimating the population QTE were proposed and discussed. To explore their relative performance in high quantile situations, a series of simulation studies are carried out considering values of $\tau$ close to 1 in weak and strong confounding settings, and  in settings with  binary, categorical or continuous exposures. Further, we generate data from location shift models with heavy tailed distributions, namely the Pareto distribution and Student's t-distribution.

As stated in Proposition \ref{pqte=cqte} and Proposition \ref{prop:ps_reg}, without interaction between exposure and confounders, the C-QTE coincides with population QTE. In our simulations, we mainly focus on this scenario by choosing a constant conditional QTE. A special case with an interaction between exposure and univariate confounder $X$ is also considered. The population QTE estimate based on PS regression is computed by empirical methods; computational details are provided in the Online Supplementary Materials. Two classic quantile regressions - a true model and a biased (`naive') model only including exposure $Z$ that ignores confounding - are also performed to further compare the results with benchmark best- and worst-case modelling scenarios. Throughout the simulations, the propensity score models are correctly specified. 

Confounding variables $\boldsymbol{X} = (X_1, X_2, ..., X_d)^\top$ are generated from either a standard normal distribution $\mathcal{N}(0,1)$ if $d=1$, or from a multivariate normal distribution  $ \mathcal{N}_d(\boldsymbol{\mu}, \boldsymbol{\Sigma})$ if $d>1$. Simulations with error $\epsilon$ arising from Pareto distribution with fixed location ($\epsilon_m=1$) and varying shape ($\theta \in \{5,7,10\}$) are presented and discussed in this section. We examine high population QTE estimates with $\tau=0.95$; results for other large $\tau$ values and data simulated from a Student's t-distribution are provided in the Online Supplementary Materials. In each simulation setting, the MSE, bias and confidence interval coverage of high population QTE estimates are computed based on 1000 simulated datasets, each with sample size 2000.

\subsection{Binary Exposure}
In a binary exposure setting, the conditional distribution of $Z$ given $\boldsymbol{X}$ follows a Bernoulli distribution with success probability $e(\boldsymbol{X})$. The true conditional QTE is chosen to be 1. The propensity scores are estimated using a correctly specified logistic regression.

\subsubsection{Univariate confounder}
\begin{figure*}[ht]
    \centering
    \includegraphics[scale=0.6]{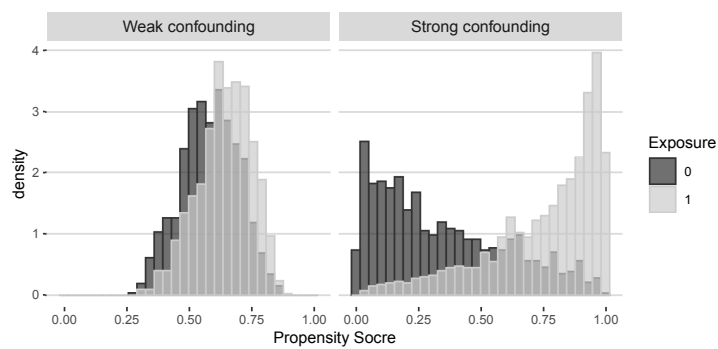}
    \caption{Distribution of the true propensity score in weak and strong univariate confounding variable settings. The light grey bars indicate those exposed ($Z=1$), the medium grey bars indicate those unexposed ($Z=0$).}
    \label{fig:1}
\end{figure*}

Let the true propensity score be given by
\begin{align*}
e(X) = 1/(1+e^{-\alpha_0-\alpha_1X}).
\end{align*}
We consider weak and strong confounding scenarios by choosing parameters $(\alpha_0, \alpha_1) = (0.5, 0.5)$ and $(\alpha_0, \alpha_1) = (0.5, 2)$, respectively. The overall proportion of receiving exposure is about 62\% in the weak confounding setting, and about 58\% in the strong confounding setting. Figure \ref{fig:1} displays the PS distributions in the two settings. More extreme propensity scores appear in the strong confounding setting such that there is less overlap between the exposed and unexposed groups. Two scenarios are considered. Without interaction, the outcome is generated according to $Y = 1 + Z + X + \epsilon$. With interaction between exposure and confounder, the outcome is generated by $Y = 1 + Z + ZX + X + \epsilon$. For the true and PS regression models, the population QTE is estimated using the empirical method detailed in the Online Supplementary Materials.

\subsubsection{Multivariate confounders}
\begin{figure*}[ht]
    \centering
    \includegraphics[scale=0.5]{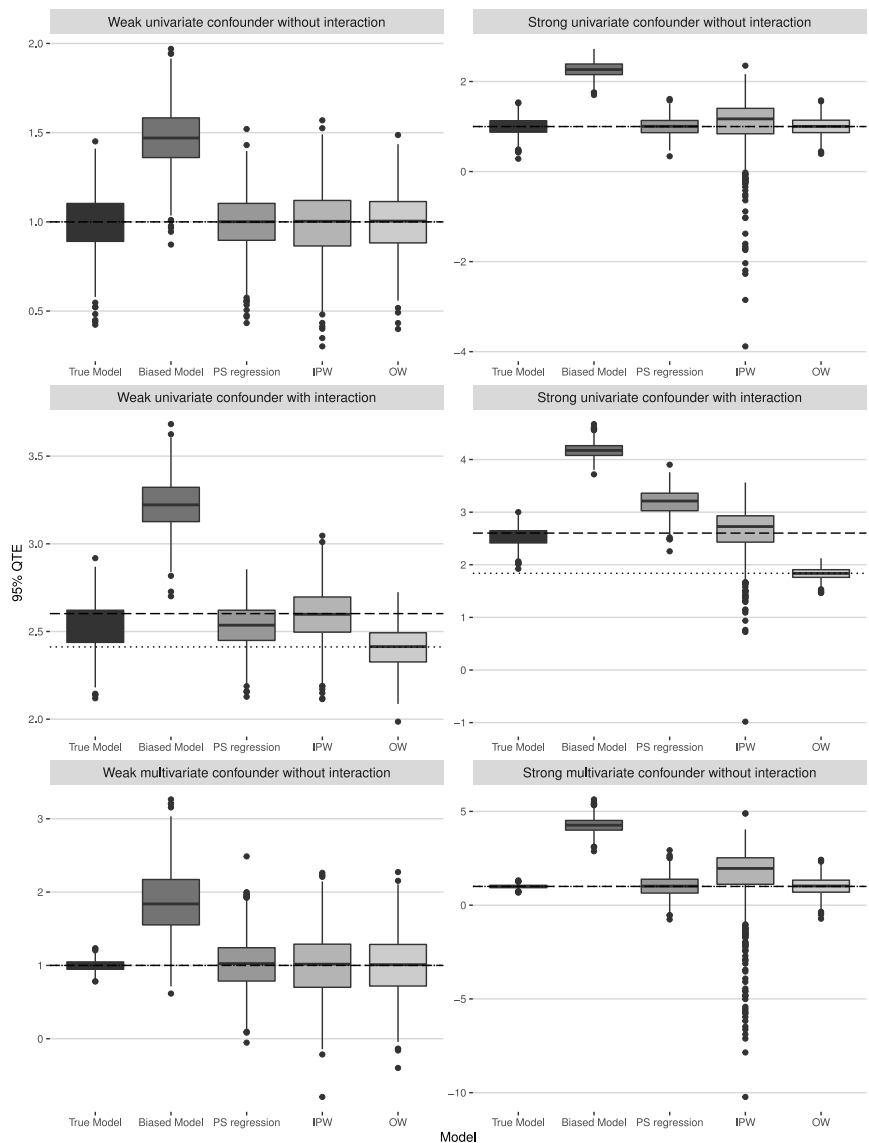}
    \caption{Bar plots of the 95\% population QTE estimates of different models in the binary exposure setting ($\epsilon_m=1$, $\theta=5$). Dashed (\dashed) denotes the true 95\% population QTE, dotted (\dotted) denotes the true 95\% WQTE. Note that the QTE and WQTE coincide in the top two and bottom two panels, where the treatment effects are homogeneous across the covariates.}
    \label{fig:2}
\end{figure*}

The confounding variables are generated according to
\begin{equation} \label{sim:eq:N4}
\begin{psmallmatrix}
X_1\\
X_2\\
X_3\\
X_4
\end{psmallmatrix} \sim \mathcal{N}_4
\begin{bmatrix}
\begin{psmallmatrix}
0\\
0\\
0\\
0
\end{psmallmatrix}\ ,\ 
\begin{psmallmatrix}
1  & 0.5 & 0.2 & 0.3 \\ 
0.5 & 1  & 0.7 & 0  \\ 
0.2 & 0.7 & 1  & 0  \\ 
0.3 & 0  & 0  & 1  \\ 
\end{psmallmatrix}
\end{bmatrix}.
\end{equation}
The true propensity score is given by $e(\boldsymbol{X}) = (1+e^{-\alpha_1X_1-\alpha_2X_2-\alpha_3X_3-\alpha_4X_4})^{-1}$.   
%\begin{align*}
%e(\boldsymbol{X}) = 1/(1+e^{-\alpha_1X_1-\alpha_2X_2-\alpha_3X_3-\alpha_4X_4}).   
%\end{align*}
We chose the parameters in the PS model to be $(\alpha_1, \alpha_2, \alpha_3, \alpha_4) = (-0.1,0.2,0.2,-0.1)$ to reflect weak confounding, and $(\alpha_1, \alpha_2, \alpha_3, \alpha_4) = (-1,2,2,-1)$ to reflect strong confounding. The overall proportion exposed is approximately 50\% both in weak and strong confounding settings. The propensity score distributions in multivariate confounding settings can be found in Figure S.1 in the Online Supplementary Materials. The outcome is generated from
\begin{align*}
    Y = 1 + Z + \sin(X_1) + X_2^2 +X_3 + X_4 + X_3X_4 + \epsilon.
\end{align*}
The barplots of the 95\% population QTE estimates with $\text{Pareto }(\epsilon_m=1,\theta=5)$ errors are presented in Figure~\ref{fig:2}, along with the true 95\% population QTE and WQTE. The MSE and absolute bias of the 95\% population QTE estimates are shown in Table \ref{Binary_MSE}. The confidence interval coverage is given in the Online Supplementary Materials (Table S.2), where the confidence intervals are computed by the rank inversion method \citep{koenker_2005, koenker1994confidence}.

Without interaction, all approaches except for the biased model (i.e., unadjusted or naive model) perform reasonably well in the weak confounding settings. Since IPW is more sensitive to extreme propensity scores, the high QTE estimates in the strong confounding setting have relatively large MSEs, large absolute bias, and poor confidence interval coverage, compared to PS regression and OW approaches. Among all the different confounding effects and dimension settings, PS regression and OW yield unbiased estimators and perform approximately equally well. 

When the data generation includes interactions between the exposure and confounders, the OW approach estimates are no longer consistent for the population QTE (see middle panels in Figure \ref{fig:2}). In contrast, the IPW estimator is  consistent for the population QTE. In the weak confounding setting, the IPW estimator is unbiased with small MSE. In the strong confounding setting, the bias and variance of the IPW estimator increase. The empirical estimator obtained from PS regression in the weak confounding setting has the smallest MSE among the three proposed approaches. However the estimator is biased, due to mis-specification of the outcome model.  In practice, computing the empirical estimates from PS regression is complex and time-consuming.

\subsection{Categorical Treatment}
In the following sections, we focus on models without any interaction between the exposure and confounders. In the categorical exposure setting, we consider $J=3$ groups. Exposure $Z$ given confounders $\boldsymbol{X}$ follows 
\begin{align*}
(D_1, D_2, D_3)\mid \boldsymbol{X} \sim \text{Multinom}(e_1(\boldsymbol{X}), e_2(\boldsymbol{X}), e_3(\boldsymbol{X})),
\end{align*}
where $D_j$ is the indicator function of treatment $Z=j$. $Z=1$ is chosen as the baseline; the pairwise exposure effect between $Z\neq 1$ and baseline is set to 1 for $Z=2$, and -1 for $Z=3$. A correctly specified multinomial logistic regression is used to estimate the propensity scores.

\subsubsection{Univariate confounder}
Let the generalized propensity score be
\begin{equation*}
\begin{aligned}
e_1(X)&=1/(1+e^{\alpha_0+\alpha_1X}+e^{\gamma_0+\gamma_1X}),\\
e_2(X)&=e^{\alpha_0+\alpha_1X}/(1+e^{\alpha_0+\alpha_1X}+e^{\gamma_0+\gamma_1X}),\\
e_3(X)&=e^{\gamma_0+\gamma_1X}/(1+e^{\alpha_0+\alpha_1X}+e^{\gamma_0+\gamma_1X}).
\end{aligned}
\end{equation*}
The parameters are chosen as $(\alpha_0, \alpha_1) = (0.5, 0.2)$, $(\gamma_0, \gamma_1) = (0.5, 0.3)$ to reflect weak confounding, and $(\alpha_0, \alpha_1) = (0.5, 1.5)$, $(\gamma_0, \gamma_1) = (0.5, 2)$ to reflect strong confounding. The overall proportion exposed to $Z=1$ was 23\%, to $Z=2$ is 39\%, and to $Z=3$ is 38\% in the weak confounding setting; and, respectively, 31\%, 31\% and 38\% in the strong confounding setting. We generate the outcome using $Y = 1 +  \mathbbm{1}(Z=2) -\mathbbm{1}(Z=3)+ X + \epsilon$.

\subsubsection{Multivariate confounders}
The confounding variables $\boldsymbol{X}=(X_1, X_2, X_3, X_4)^\top$ are generated from the same distribution as specified in \eqref{sim:eq:N4}. The true generalized propensity score has form
\begin{equation*}
\begin{aligned}
e_1(X)&=1/(1+e^{\boldsymbol{X}^\top\boldsymbol{\alpha}}+e^{\boldsymbol{X}^\top\boldsymbol{\gamma}}),\\
e_2(X)&=e^{\boldsymbol{X}^\top\boldsymbol{\alpha}}/(1+e^{\boldsymbol{X}^\top\boldsymbol{\alpha}}+e^{\boldsymbol{X}^\top\boldsymbol{\gamma}}),\\
e_3(X)&=e^{\boldsymbol{X}^\top\boldsymbol{\gamma}}/(1+e^{\boldsymbol{X}^\top\boldsymbol{\alpha}}+e^{\boldsymbol{X}^\top\boldsymbol{\gamma}}).
\end{aligned}
\end{equation*}
The parameters $\boldsymbol{\alpha} = (\alpha_1, \alpha_2, \alpha_3, \alpha_4)^\top = \eta\times(0.1,0.2,0.2,0.1)$, $\boldsymbol{\gamma} = (\gamma_1, \gamma_2, \gamma_3, \gamma_4) = \eta\times (0.1,0.1,0.2,0.2)$. $\eta$ is set with values 1 and 5 to reflect weak and strong confounding, respectively. The outcome is generated according to $Y = 1 +  \mathbbm{1}(Z=2) -\mathbbm{1}(Z=3)+ \sin(X_1) + X_2^2 +X_3 + X_4 + X_3X_4 + \epsilon$.

\begin{figure*}[ht]
    \centering
    \includegraphics[scale=0.5]{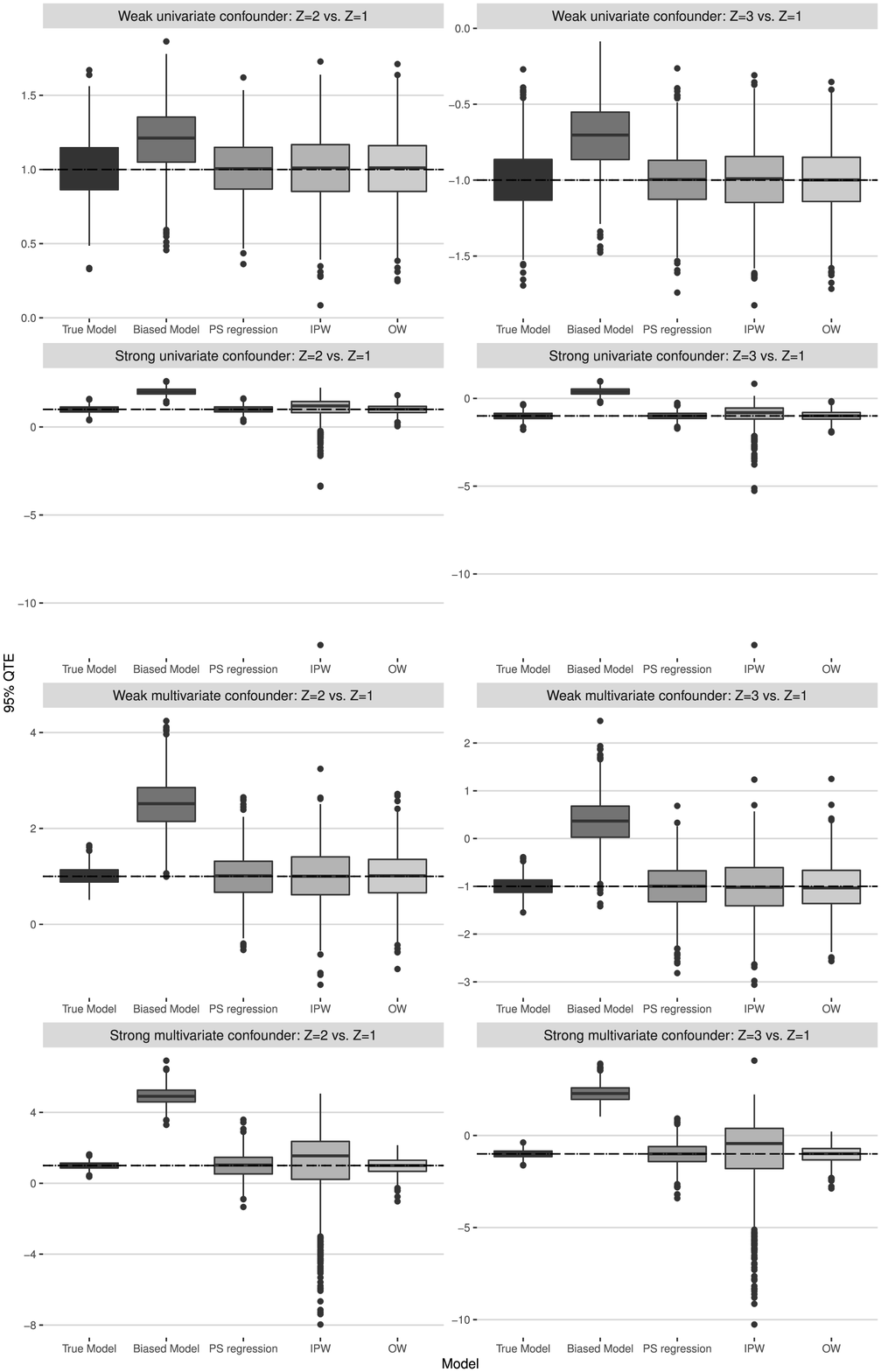}
    \caption{Bar plots of the 95\% pairwise population QTE estimates between $Z=2$, $Z=3$ versus baseline $Z=1$ of different models in the categorical exposure setting ($\epsilon_m=1$, $\theta=5$). Dashed (\dashed) denotes the true 95\% population QTE, dotted (\dotted) denotes the true 95\% WQTE. Note that the QTE and WQTE coincide in this setting, where the treatment effects are homogeneous across the covariates.}
    \label{fig:3}
\end{figure*}

The barplots of the 95\% pairwise population QTE estimates with $\epsilon_m=1$ and $\theta=5$ are displayed in Figure~\ref{fig:3}. The associated MSE and absolute value of bias of the QTE estimates are listed in Table \ref{Categorical_MSE_Z23}. Confidence interval coverage is summarized in Table S.5 in the Online Supplementary Materials. As expected, all the three models (PS regression, IPW, and OW) perform well in the weak univariate confounding setting, though IPW tends to have slightly larger MSE than the other two approaches. Under strong confounding, IPW estimators are biased with larger MSE than PS regression and OW, indicating IPW is not robust in strong confounding, likely due to near violations of positivity in finite samples. This type of behavior has also been observed in the context of the ATEs (see \cite{li2018addressing, li2019propensity}). Comparing PS regression to OW in this scenario, the latter exhibits smaller MSE and slightly larger bias than the former. Table S.5 in the Online Supplementary Materials provides more evidence to support these conclusions. The coverage of confidence intervals based on the IPW estimator decreases dramatically in strong confounding settings.

\subsection{Continuous Treatment}
In continuous exposure setting, $Z$ is generated from a Normal distribution. As stated previously, weights derived from the generalized propensity score may be volatile. Therefore, we bin the continuous exposure into 10 groups according to its empirical quantiles, and treat these as a categorical variable with 10 levels. We use the true underlying continuous distribution of $Z$ with CDF $F$ to compute the propensity scores. A simple linear regression is used to estimate the mean and standard deviation of $Z$.

\subsubsection{Univariate confounder}
We generate continuous $Z$ from $\mathcal{N}(a_0+a_1X, 5)$. The parameters $(a_0, a_1)$ are set to be $(1,0.3)$ and $(1, 3)$ to generate weak and strong confounding, respectively. The outcome is generated according to $Y = 1 + Z + X + \epsilon$.

\subsubsection{Multivariate confounders}
In the multivariate confounding  scenario, the continuous exposure doses follow a Normal distribution given by $\mathcal{N}(a_1X_1+a_2X_2+a_3X_3+a_4X_4, 5)$, and the confounding variables $\boldsymbol{X} = (X_1,...,X_4)^\top$ are drawn from the multivariate Normal distribution given in (\ref{sim:eq:N4}). We set $(a_1, a_2, a_3, a_4) = 0.1 \times(1,2,2,1)$ to generate weak confounding, and $(a_1, a_2, a_3, a_4)=0.8\times(1,2,2,1)$ to generate strong confounding. The outcome is generated as
\begin{align*}
Y = 1 + Z + \text{sin}(X_1) + X_2^2 + X_3 + X_4 + X_3X_4 + \epsilon.    
\end{align*}

Simulation results are summarized in Table S.8 in the Online Supplementary Materials. The same patterns as are observed in the binary exposure and categorical exposure settings persist in the continuous exposure setting. The three proposed approaches perform approximately equally well in the weak confounding scenario, while PS regression and OW outperform IPW in the strong confounding scenario. In addition, 99\% QTE estimates with Pareto distributed errors, and 95\% QTE estimates with Student's t-distributed errors are provided in the Online Supplementary Materials. Again, the same patterns of results persist.

\section{High population QTE in the Bavarian Danube River}\label{example}
Eutrophication and metal pollution are two major environmental problems in fluvial ecological systems. Excessive input of phosphorus is closely related to eutrophication \citep{johnson1997landscape}. Agricultural activity such as the use of fertilizers is one of the major sources of phosphorus to aquatic ecosystems \citep{carpenter1998nonpoint, ekholm2000relationship}. In addition, human activities also lead to an increase in concentration of heavy metals such as copper in aquatic systems \citep{andrade2004effects}, which may result in heavy metal poisoning. For example, chronic excessive copper exposure can cause liver and kidneys damage in humans. The U.S.~Environmental Protection Agency has established a recommended upper limit for copper in public water systems of 1.3 mg/L  \cite{NAP9782}. According to \cite{food2001dietary}, the Tolerable Upper Intake Level (UL) for adults is 10 mg/day, a value chosen to protect against critical adverse effects such as liver damage. Though metal pollution is often linked to eutrophication in aquatic ecosystems \citep{lopez2003comparison}, the relationship between heavy metal pollution and eutrophication is not well understood. It has been shown that periphyton, a complex mixture of algae and other microbes, accumulates metals from surrounding water and sediments and that the composition of species is affected by metal pollution \citep{tang2014heavy,morin2008long}, but the impact of heavy metals may be modulated by other elements such as phosphorus \citep{serra2010influence}. If the presence of phosphorus increases the concentration of copper, affected bodies of water may suffer the double impact of increased algal bloom along with a bioaccumulation of toxic metals. Understanding the impact of phosphorus, an important driver of eutrophication, and heavy metal pollution could provide additional incentives for tackling this important environmental concern.

We illustrate the application of high causal QTE estimation using data taken from the catchment of the Bavarian Danube River. The data are provided by the Bavarian Environmental Agency (\url{www.gkd.bayern.de}). Daily measurements of meteorology, chemistry, and river- and lake-related variables were collected. The daily series have lengths ranging from several months to 36 years, depending on the  program through which they were collected. We assessed the 95\% population QTE of phosphorous content on copper concentration in the river controlling for the potentially confounding effect of precipitation (since heavy rain is likely to be related to high phosphorous concentration).
\begin{figure}[ht]
    \centering
    \includegraphics[scale=0.7]{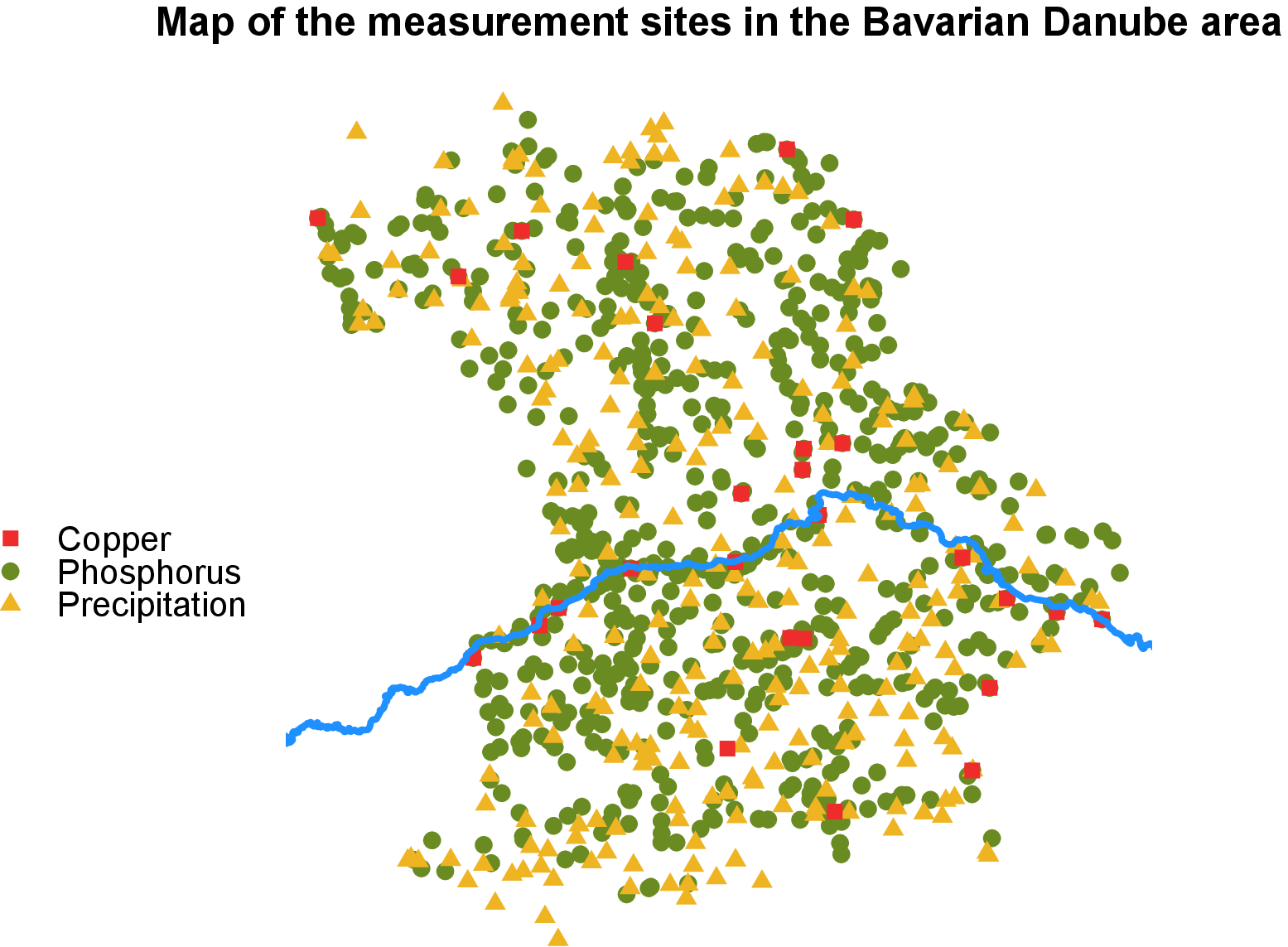}
    \caption{Map of all the measurement sites in the Bavarian Danube river area.}
    \label{fig:4}
\end{figure}

In order to estimate the QTE, we assume the data satisfy the assumptions listed in Section 2 above. Additional assumptions are also required. First, since meteorology and chemistry variables were measured in different locations, we matched each copper measurement site with the nearest phosphorus site and the nearest precipitation site based on Euclidean distance, making the assumption that these sites' data values were similar. Further, we wished to ensure (approximate) independence of errors in our analyses. Thus, to address the temporal and spatial correlation, we retrieved the monthly maximum copper concentration; proceding this way did not reduce the number of measures of data too greatly since the vast majority of the sites measured copper concentration only once a month. We also randomly drew 50\% of the data, thereby breaking the strict spatial and temporal relationship between the measures in the analytic sample. In total, 25 copper measurement sites were randomly selected; our random selection of monthly aggregate data resulted in an analytic sample of 1,099 observations. Table \ref{tab:x_dist_for_z} displays the means and SDs of log-transformed phosphorus (mg/L) and copper concentration (mg/L) in the river, categorized by precipitation (mm). Scatterplots of log-transformed phosphorus and log-transformed copper are shown in Figure~S.2 in the Online Supplementary Materials.

We log-transformed the continuous exposure phosphorous concentration, as well as the outcome, copper concentration. After controlling for precipitation, the estimated causal effect of phosphorus on the 95th quantile of the copper concentration distribution (95\% population QTE) in the Danube River were computed; the resulting estimates are  summarized in Table \ref{tab:ex:copper}. The three proposed methods provide similar results. In the 95th quantile of the distribution of copper, every 10-fold increase in phosphorus will increase copper concentration in the river by nearly two-fold ($\exp(0.63)\approx 1.88$), unconfounded by any effect of precipitation.

We performed a second analysis, to determine the causal effect of precipitation on the 95th quantile of the total phosphorous concentration distribution in the river, adjusting for the potential confounding effect of water temperature. Warmer temperatures cause more evaporation, and may increase the frequency or intensity of rainfall. Results demonstrate that precipitation has a positive effect on the upper tail of the phosphorus concentration distribution. Estimates, SEs and CIs are provided in the Online Supplementary Materials. 

Our analysis of the Bavarian Danube data has some limitations. Matching of precipitation and phosphorus measurements to copper measurement sites was based on Euclidean distance, but did not account for topology such as mountains, hills, or other geographical features. Further, we used simple random sampling to detrend the data and attempt to break any spatio-temporal correlation. However, more sophisticated approaches could have been used (see \cite{asadi2015extremes,engelke2020graphical,mhalla2020causal}). 

\section{Conclusion} \label{conclusion}
In this article, we use the population QTE and define a novel estimator --- the WQTE --- of which the population QTE is a special case, using a previously defined class of balancing weights developed for estimating average effects; these weights rely on functions of the PS to weight each group to a selected target population, allowing comparisons of the distributional treatment effect over a subpopulation of interest. Large sample asymptotic properties of the WQTE estimators were derived. In particular, we have shown how to estimate distributional (quantile) causal effects by a simple two-step approach.In causal inference, the population QTE over a target population may be of primary interest rather than the conditional QTE (C-QTE). The population QTE is not easy to recover from the C-QTE. In this article, we have demonstrated that for a linear causal quantile model, the population QTE coincides with C-QTE for causal quantile models without interactions between the treatment and the covariates (i.e., a homogeneous C-QTE) and have showed that they differ when the C-QTE is heterogeneous. Three methods -- PS regression, IPW and OW -- were considered. We have proven that the WQTE estimator with IPW is consistent for population QTE, whether treatment effects are homogeneous or heterogeneous. In addition, we have demonstrated that the WQTE estimator with OW is consistent for the population QTE if the C-QTE is homogeneous. Further, practical parametric estimators were presented which could accommodate an exposure that is of any form (binary, discrete or continuous).   

The proposed estimator and developed asymptotic theory are valid for any fixed quantile, when sample size goes to infinity. Considering the data sparsity challenges of quantile regression in high quantiles, we studied the finite sample behavior of the regression and weighting estimators in high quantile settings across a series of simulation studies to examine their consistence. We conclude that in practice, if researchers believe the C-QTE is homogeneous (i.e., no interactions between exposure and confounders), PS regression and OW are to be preferred since both estimators are unbiased with small variance. If, however, the conditional QTE is more likely to be heterogeneous, IPW is  recommended. Although IPW is sensitive to extreme PS values, it is both simple to compute (unlike the PS regression estimator) and consistent (unlike OW). While our simulation results further support the consistency of estimators even in relatively high quantiles (e.g., 95th and 99th quantiles), estimation in extreme quantiles requires extrapolation techniques and extreme value theory -- a topic that we are tackling in subsequent work.

In a continuous exposure setting, we opted to bin the exposure and treated it as a discrete variable for the purpose of weight construction, as the generalized PS is often unstable. An interesting future avenue of consideration would be to consider the generalized propensity score for the continuous exposure, and investigate relaxations of the positivity assumption to allow for positivity within set ``distances'' of the observed exposure level. Another consideration is that of complex exposure distributions. For instance, precipitation often exhibits a mixture distribution, with a large spike of zeros. A more sophisticated consideration of such an exposure distribution warrants further investigation. 

\section*{Acknowledgments}
This work was supported by Discovery Grants from the Natural Sciences and Engineering Research Council (NSERC) of Canada (grants RGPIN-2019-04230 to E.E.M.M. and RGPIN-2015–06801 to J.G.N.), the Fonds de Recherche du Qu\'ebec, Sant\'e (award FRQS-S CB Sr 34840 to E.E.M.M.) and the Institut de Valorisation des Donn\'ees (IVADO) Grants (grants PRF-2019-7771647733). 

\section*{Supporting Information}
Additional information and supporting material for this article is available online at the journal's website.

\newpage
\bibliography{biblio.bib}
\bibliographystyle{apalike}

%======== Tables after the reference list ========

\begin{sidewaystable*}[ht]
\centering 
\begin{threeparttable}
\caption{MSE and absolute bias of the 95\% population QTE estimates in the binary exposure setting. The dimension of $\boldsymbol{X}$ is $d\geq 1$.} 
\label{Binary_MSE} 

\begin{tabular}{ll ccc | ccc | ccc}
\toprule
&\multirow{2}{*}{Model} & \multicolumn{3}{c}{$d=1$ without interaction} & \multicolumn{3}{c}{$d=1$ with interaction} &\multicolumn{3}{c}{$d=4$ without interaction}  \\ 
\cmidrule(l){3-5} \cmidrule(l){6-8} \cmidrule(l){9-11}
&& \mc{$\theta=5$} & \mc{$\theta=7$} & \mc{$\theta=10$} & \mc{$\theta=5$} & \mc{$\theta=7$} & \mc{$\theta=10$}& \mc{$\theta=5$} & \mc{$\theta=7$} & \mc{$\theta=10$}\\

\midrule
&& \multicolumn{9}{c}{\textit{Weak Confounding Setting}}\\\addlinespace
\multirow{5}{*}{MSE}&True model&0.01&0.00&0.00&0.02&0.01&0.00&0.00&0.00&0.00\\ 
&Biased model &0.23&0.24&0.23&0.41&0.40&0.40&0.94&1.00&0.96\\
&PS regression &0.01&0.00&0.00&0.02&0.01&0.00&0.12&0.12&0.12\\
&IPW &0.02&0.01&0.01&0.02&0.02&0.02&0.19&0.19&0.18\\
&OW &0.01&0.01&0.01&0.05&0.05&0.05&0.17&0.17&0.17\\\addlinespace
\midrule
\multirow{5}{*}{$\mid$Bias$\mid$}&True model&0.00&0.00&0.00 &0.07&0.02&0.01&0.00&0.00&0.00\\ 
&Biased model &0.47&0.48&0.47&0.62&0.62&0.61&0.87&0.90&0.87\\
&PS regression &0.01&0.00&0.00&0.07&0.02&0.01&0.02&0.00&0.01\\
&IPW &0.01&0.00&0.00&0.01&0.00&0.00&0.00&0.02&0.01\\
&OW &0.00&0.00&0.00&0.19&0.19&0.19&0.00&0.02&0.01\\\addlinespace
\midrule
&& \multicolumn{9}{c}{\textit{Strong Confounding Setting}}\\\addlinespace
\multirow{5}{*}{MSE}&True model &0.01&0.00&0.00&0.03&0.01&0.00&0.01&0.00&0.00\\ 
&Biased model &1.77&1.79&1.79&2.48&2.52&2.52&10.86&11.01&10.95\\
&PS regression &0.01&0.00&0.00&0.40&0.59&0.72&0.30&0.29&0.30\\
&IPW &0.17&0.15&0.15&0.18&0.16&0.15&3.06&3.04&3.21\\
&OW &0.01&0.01&0.01&0.61&0.59&0.57&0.21&0.20&0.21\\\addlinespace
\midrule
\multirow{5}{*}{$\mid$Bias$\mid$}&True model &0.00&0.00&0.00&0.07&0.02&0.01&0.01&0.00&0.00\\ 
&Biased model &1.33&1.34&1.34&1.57&1.58&1.58&3.27&3.29&3.28\\
&PS regression &0.01&0.01&0.00&0.59&0.75&0.84&0.01&0.00&0.01\\
&IPW &0.05&0.04&0.02&0.04&0.03&0.01&0.53&0.50&0.47\\
&OW &0.00&0.00&0.00&0.77&0.76&0.75&0.01&0.02&0.01\\\addlinespace
\bottomrule 
\end{tabular} 
\begin{tablenotes}
%\footnotesize
\item NOTE: Error follows Pareto distribution with location $\epsilon_m=1$ and shape $\theta$. The ``biased model'' is the naive model only including exposure $Z$ that ignores confounding $\boldsymbol{X}$.
\item[$\dagger$] Abbreviations: PS, propensity score; IPW, inverse probability weighting; OW, overlap weighting.
\end{tablenotes}
\end{threeparttable}
\end{sidewaystable*}

\begin{sidewaystable*}[ht]
\centering 
\begin{threeparttable}
\caption{MSE and absolute bias of the 95\% pairwise population QTE estimates between $Z=2$, $Z=3$ versus baseline $Z=1$ in the categorical exposure setting. The dimension of $\boldsymbol{X}$ is $d\geq 1$.} 
\label{Categorical_MSE_Z23} 

\begin{tabular}{ll ccc ccc | ccc ccc}
\toprule
&\multirow{3}{*}{Model} & \multicolumn{6}{c}{$d=1$ without interaction} & \multicolumn{6}{c}{$d=4$ without interaction}  \\ 
\cmidrule(l){3-8} \cmidrule(l){9-14} 
& & \multicolumn{3}{c}{$Z=2$ vs. $Z=1$} & \multicolumn{3}{c}{$Z=3$ vs. $Z=1$}& \multicolumn{3}{c}{$Z=2$ vs. $Z=1$} & \multicolumn{3}{c}{$Z=3$ vs. $Z=1$}\\ \cmidrule(l){3-5} \cmidrule(l){6-8} \cmidrule(l){9-11} \cmidrule(l){12-14}
& & \mc{$\theta=5$} & \mc{$\theta=7$} & \mc{$\theta=10$} & \mc{$\theta=5$} & \mc{$\theta=7$} & \mc{$\theta=10$}& \mc{$\theta=5$} & \mc{$\theta=7$} & \mc{$\theta=10$} & \mc{$\theta=5$} & \mc{$\theta=7$} & \mc{$\theta=10$}\\

\midrule
&& \multicolumn{12}{c}{\textit{Weak Confounding Setting}}\\\addlinespace
\multirow{5}{*}{MSE}&True model &0.01&0.00&0.00&0.01&0.00&0.00 &0.01&0.00&0.00&0.01&0.00&0.00\\ 
&Biased model &0.06&0.06&0.05&0.10&0.10&0.10&2.92&2.93&2.90&2.47&2.34&2.39\\
&PS regression &0.01&0.00&0.00&0.01&0.00&0.00&0.21&0.22&0.21&0.23&0.21&0.22\\
&IPW &0.02&0.02&0.02&0.02&0.02&0.02&0.34&0.35&0.36&0.37&0.34&0.38\\
&OW &0.02&0.02&0.01&0.02&0.01&0.01&0.26&0.27&0.28&0.28&0.25&0.29\\\addlinespace
\multirow{5}{*}{$\mid$Bias$\mid$}&True model &0.00&0.00&0.00&0.00&0.00&0.00&0.00&0.00&0.00&0.00&0.00&0.00\\ 
&Biased model &0.19&0.20&0.19&0.29&0.30&0.30&1.63&1.63&1.62&1.48&1.44&1.45\\
&PS regression &0.00&0.00&0.00&0.00&0.00&0.00&0.02&0.01&0.00&0.02&0.00&0.00\\
&IPW &0.00&0.00&0.00&0.00&0.00&0.00&0.01&0.00&0.01&0.01&0.02&0.01\\
&OW &0.00&0.00&0.00&0.00&0.00&0.00&0.02&0.01&0.02&0.02&0.03&0.01\\\addlinespace

&& \multicolumn{12}{c}{\textit{Strong Confounding Setting}}\\\addlinespace
\multirow{5}{*}{MSE}&True model &0.01&0.00&0.00&0.01&0.00&0.00&0.01&0.00&0.00&0.01&0.00&0.00\\ 
&Biased model &1.12&1.14&1.14&2.12&2.14&2.14&18.00&18.20&18.40&12.61&12.46&12.69\\
&PS regression &0.01&0.00&0.00&0.01&0.00&0.00&0.52&0.49&0.53&0.42&0.40&0.42\\
&IPW &0.22&0.17&0.18&0.22&0.16&0.17&3.93&4.31&3.91&4.03&4.31&4.03\\
&OW &0.02&0.02&0.01&0.02&0.01&0.01&0.19&0.21&0.20&0.18&0.19&0.19\\\addlinespace
\multirow{5}{*}{$\mid$Bias$\mid$}&True model &0.00&0.00&0.00&0.00&0.00&0.00&0.00&0.00&0.00&0.01&0.00&0.00\\ 
&Biased model &1.05&1.06&1.06&1.45&1.46&1.46&4.21&4.24&4.26&3.52&3.50&3.53\\
&PS regression &0.00&0.00&0.00&0.00&0.00&0.00&0.00&0.01&0.01&0.02&0.00&0.03\\
&IPW &0.06&0.05&0.05&0.06&0.05&0.06&0.07&0.07&0.03&0.12&0.07&0.09\\
&OW &0.00&0.00&0.00&0.00&0.00&0.00&0.02&0.03&0.03&0.01&0.03&0.03\\\addlinespace
\bottomrule 
\end{tabular} 
\begin{tablenotes}
\item NOTE: Error follows Pareto distribution with location $\epsilon_m=1$ and shape $\theta$. The ``biased model'' is the naive model only including exposure $Z$ that ignores confounding $\boldsymbol{X}$.
\item[$\dagger$] Abbreviations: PS, propensity score; IPW, inverse probability weighting; OW, overlap weighting.
\end{tablenotes}
\end{threeparttable}
\end{sidewaystable*}

\begin{table*}[ht]
    \centering
    \begin{threeparttable}
    \caption{Mean (SD) of log-transformed phosphorus (mg/L) and log-transformed copper (mg/L).}
    \label{tab:x_dist_for_z}
    \begin{tabular}{l ccc}
    \toprule
    \multirow{2}{*}{Log-transformed concentration} &  No rain & Light/Moderate & Heavy\\
    & ($n=590$) & ($n=454$) & ($n=55$)\\
    \midrule
    phosphorus & -2.26 (0.64) & -2.20 (0.72) & -2.16 (0.85)\\
    copper & -5.80 (0.69) & -5.71 (0.71) & -5.58(0.80)\\
    \bottomrule
    \end{tabular}
    \end{threeparttable}
\end{table*}

\begin{table*}[ht]
\centering
\begin{threeparttable}
\caption{95\% population QTE estimates, SEs and 95\% CIs associated with a 10-fold increase in phosphorous on total log-copper concentration (mg/L) in Danube river.}
\label{tab:ex:copper}
    \begin{tabular}{l ccc}
    \toprule
      &  PS regression & IPW & OW\\
    \midrule
    Estimate  & 0.631  &0.627  & 0.627\\
    Std. Error &0.100 & 0.116 & 0.116 \\
    95\% CI & (0.310, 0.747) & (0.398,  0.727) & (0.399,  0.727)\\
    \bottomrule
    \end{tabular}
\begin{tablenotes}
\item NOTE: Confidence intervals are computed by the rank inversion method described by \cite{koenker1994confidence}.
\end{tablenotes}
\end{threeparttable}
\end{table*}

\end{document}